\documentclass[12pt,reqno]{article}
\usepackage{amsmath, amsthm, amssymb,epsfig,alltt}

\begin{document}

%%GO%%%%%%%%%%%%%%%%%%%%%%%%%%%%%%%%%%%%%%%%%%%%%%%%%%%%%%%%%%%%%%%%

\title{ Stretching the electron as far as it will go }

\author{  Gordon W. Semenoff\footnote{This work is supported in part by
NSERC of Canada.}
\\{\small
Department of Physics and Astronomy}\\{\small University of
British Columbia,  Vancouver, British Columbia, Canada V6T 1Z1}
\\~~\\Pasquale Sodano\thanks{  Dipartimento di Fisica e Sezione INFN,
 Università degli Studi di Perugia,
 Via A. Pascoli, 06123, Perugia}\\
{\small Progetto Lagrange, Fondazione C.R.T. e Fondazione I.S.I.}\\
{\small c/o Dipartimento di Fisica- Politecnico di Torino,}\\
{\small Corso Duca degli Abbruzzi 24 Torino-Italy} }

\maketitle                 % Produces the title.

\abstract{Effects associated with the existence of isolated zero
modes of Majorana fermions are discussed.   It is argued that the
quantization of this system necessarily contains highly extended
quantum states and that populating and depopulating such states by
interacting with the quantum system leads to long-ranged
teleportation-like processes.   Also leads to spontaneous
violation of fermion parity symmetry.   A quasi-realistic model
consisting of a quantum wire embedded in a p-wave superconductor
is discussed as an explicit example of a physical system with an
isolated Majorana zero mode.  }

\vskip 2cm

\section{Introduction}

It is a great pleasure to dedicate this article to the 100'th
anniversary of the birth of Ettore Majorana.  As a testimony to
his lasting influence on science, we shall describe how one of his
great insights, used in a modern context, can be related to a
particular macroscopic quantum phenomenon.

%Due to the entanglement, of majorana fermions as opposed to dirac
%fermions. which we shall explain.

The idea is related to the observation by Majorana that a
relativistic fermion such as the electron can be meaningfully
decomposed into more basic degrees of freedom, essentially by
taking the real and imaginary parts of its
wave-function~\cite{Major}.  In relativistic field theory, what
one obtains are called Majorana fermions, which have become the
basic building blocks of supersymmetric field theories and supply
a scenario whereby the neutrinos which are observed in nature can
have mass.  We shall use this idea in a different context.  In
quantum condensed matter, the appearance of an emergent Majorana
fermion would provide an excitation of a system that has minimal
degrees of freedom. The wave-function of the single-particle state
would obey a Majorana condition, which would forbid quantum
fluctuations of its phase.  The utility of this fact has already
been recognized in the context of quantum
computing~\cite{kitaev}-\cite{dorner}. In the present manuscript,
we will elaborate on our previous observation \cite{semsod} that
in some cases this can provide isolated states with wave-functions
which are peaked at multiple, well separated locations.  In a
controlled setting, this can be used to create a condensed matter
realization of the Einstein-Podolsky-Rosen effect and even a
version of teleportation by long-ranged tunnelling.

Majorana's original motivation for inventing the Majorana fermion
was to avoid the negative energy states that relativistic
particles invariably seem to possess by identifying the negative
and positive energy states of a relativistic wave equation as
manifestations of the same quantum excitation.

In second quantization, the positive energy state can be occupied
by a particle. Filling a positive energy state creates an excited
state of the system with positive energy.  On the other hand, a
negative energy state should be regarded as typically being
already filled by a particle. An excitation of the system is then
found by emptying the negative energy state, or creating a hole.
The system is put in a higher energy state by removing a negative
energy particle, equivalently, creating a positive energy hole.

Majorana's idea can be implemented when there is a particle-hole
symmetry. Then, for a given particle state, there exists a hole
state with the same energy and with a wave-function that is
related to the particle wave-function by a simple transformation.
Then, by making the appropriate identification, one could indeed
identify these as one and the same quantum state. Of course, the
resulting system has half as many degrees of freedom.\footnote{For
a comprehensive account of issues to do with positive and negative
energy modes of relativistic bosons and fermions, see the series
of papers~\cite{Nielsen:1998mc}-\cite{Habara:2005ka}.}

To illustrate the idea, let us recall the conventional second
quantization of complex fermions, which could be either
relativistic or non-relativistic.  We begin with the assumption
that in some approximation it makes sense to discuss a single
non-interacting particle whose wave-function obeys the
Schr\"odinger equation
\begin{equation}\label{dirac0}
i\hbar \frac{\partial}{\partial t}\Psi(\vec x,t) = H_0\Psi(\vec
x,t)
\end{equation}
where $H_0$ is the single-particle Hamiltonian operator.
Generally, as in the case of the Dirac equation, the Hamiltonian
$H_0$ could be a matrix, as well as a differential operator, and
$\Psi(\vec x,t)$ a column vector whose indices we shall
suppress.\footnote{An example is the Dirac Hamiltonian in
3+1-dimensions $$ H_0=i\vec\alpha\cdot\vec\nabla + \beta m$$ where
$\vec\alpha$ and $\beta$ are a set of four Hermitian,
anti-commuting $4\times4$ Dirac matrices. There exists a matrix
$\Gamma$ with the property $\Gamma \vec\alpha~\Gamma=\vec
\alpha^*$ and $\Gamma\beta\Gamma=-\beta^*$, so that, $\Gamma
H_0\Gamma = -H_0^*$ and $\Gamma\psi_E^*=\psi_{-E}$.  This is a
one-to-one mapping of positive to negative energy states.
Explicitly, if the matrices are represented by $\vec\alpha =
\left(\begin{matrix}\vec\sigma & 0\cr
0&-\vec\sigma\end{matrix}\right)$  with $\vec\sigma$ the Pauli
matrices and $\beta=\left(\begin{matrix} 0 & 1\cr
1&0\cr\end{matrix}\right)$, then we can form the matrix
$\Gamma=\left(
\begin{matrix} 0&-i\sigma^2\cr i\sigma^2 &
0\cr\end{matrix}\right)$.  Note that, in this case
$\Gamma=\Gamma^*$ and $\Gamma^2=1$.  A Majorana fermion obeys the
reality condition $\Psi(\vec x,t)=\Gamma\Psi^*(\vec x,t)$.} The
second-quantized field operator also typically obeys this wave
equation plus the equal-time anti-commutation relation
\begin{equation} \left\{ \Psi(\vec x,t),\Psi^\dagger(\vec
y,t)\right\}=\delta(\vec x - \vec y)\label{cacr}\end{equation} It
is this anti-commutator which defines $\Psi(x,t)$ as an operator.
It can further be used to derive the wave equation (\ref{dirac0})
from the second quantized Hamiltonian,
\begin{equation}
H=\int dx :\Psi^\dagger(x,t)H_0\Psi(x,t):
\end{equation}
using the Hamilton equation of motion
$$
i\hbar \frac{\partial}{\partial t}\Psi(x,t)=\left[
\Psi(x,t),H\right]
$$

 We shall assume that $H_0$ is a Hermitian operator which  has eigenfunctions
 and a spectrum of real eigenvalues
$$
H_0\psi_E(x)=E\psi_E(x)
$$
The energy $E$ can be both positive and negative, in fact for the
relativistic electron, if (\ref{dirac0}) were the Dirac equation,
there are necessarily negative eigenvalues and the spectrum is
unbounded below. The eigenfunctions obey the orthogonality and
completeness relations
\begin{equation}
\int d\vec x \psi^{\dagger}_E(\vec x)\psi_{E'}(\vec x)=
\delta_{EE'}~~,~~\sum_E\psi_E(\vec x)\psi_E^\dagger(\vec
y)=\delta(\vec x - \vec y)\label{comp}\end{equation} The delta
function and summation in these formulae should be understood in a
generalized sense where they are a Kronecker delta and a sum for
discrete components of the spectrum and a Dirac delta function and
integral for continuum spectrum.

 In this system, one then forms the second quantized field operator by
superposing the wave-functions with creation and annihilation
operators,
$$
\Psi(x,t)= \sum_{E>0}\psi_E(x)e^{-iEt/\hbar}a_E +
\sum_{E<0}\psi_E(x)e^{-iEt/\hbar}b_{-E}^\dagger
$$
Here, $a_E$ is the annihilation operator for a particle with
energy $E$ and $b_{-E}^\dagger$ is the creation operator for a
hole with energy $-E$.  When they obey the algebra with
non-vanishing anti-commutators
$$
\left\{ a_E, a^\dagger_{E'}\right\}=\delta_{EE'}~~,~~ \left\{
b_{-E}, b^\dagger_{-E'}\right\}=\delta_{EE'}
$$
the $\Psi(\vec x,t)$ obeys the anticommutator (\ref{cacr}). The
completeness condition in Eq.~(\ref{comp}) is essential for
establishing this.

The ground state of the system, $|0>$, is the state where all
positive energy levels are empty and where all negative energy
levels are filled, or alternatively all hole states are empty.
 In the second quantized language, it is annihilated by the
 annihilation operators,
$$
a_E|0> = 0 = b_{-E}|0>
$$
Excited states are created by operating on $|0>$ with
$a_E^\dagger$ and $b_{-E}^\dagger$.  The excitations created by
$a_E^\dagger$ are particles, those created by $b^\dagger_{-E}$ are
anti-particles, or holes.  A typical state is
$$
a_{E_1}^\dagger\ldots a_{E_m}^\dagger b_{E_1}^\dagger\ldots
b_{E_n}^\dagger |0>
$$
and such states form a basis for the Fock space of the second
quantized theory.

One can formulate Majorana fermions for a system of this kind if
there exists a particle-hole symmetry, or, in the relativistic
context, a charge conjugation symmetry. For example, consider the
situation where a matrix $\Gamma$ exists such that, for
eigenstates of $H_0$,
\begin{equation}\label{particle-holesymmetry}
\psi_{-E}(x)=\Gamma\psi^*_{E}(x)
\end{equation}
(This implies that $\Gamma^*\Gamma=1=\Gamma\Gamma^*$.) Then, the
particles and holes have identical spectra. A Majorana fermion is
formed by treating the particle and hole with the same energy as a
single excitation. The second quantized field operator is
$$
\Phi(x,t)= \sum_{E>0}\left( \psi_E(x)e^{-iEt/\hbar}a_E +
\Gamma\psi^*_E(x)e^{iEt/\hbar}a_{E}^\dagger\right)
$$
This fermion does not have both particles and anti-particles. The
ground state $|0>$ is annihilated by $a_E$ $$ a_E|0>=0~~~\forall
a_E $$
 and $a_E^\dagger$ creates particles, so that the excited states of the system
 are
 $$
 a_{E_1}^\dagger a_{E_2}^\dagger... a_{E_k}^\dagger |0>
 $$
 The field operator is
(pseudo-)real in the sense that it obeys
\begin{equation}
\label{maj} \Phi(x,t) = \Gamma \Phi^*(x,t)
\end{equation}
It obeys the anti-commutation relation
\begin{equation}
 \left\{ \Phi(\vec x,t), \Phi^\dagger(\vec
y,t)\right\}= \delta(\vec x-\vec y)
\end{equation}

To be concrete, in a system of complex fermions where the
Hamiltonian such that the spectrum has the particle-hole symmetry
(\ref{particle-holesymmetry}),  we could decompose the complex
fermion into two Majorana fermions by taking the real and
imaginary parts,
$$
\Phi_1(x,t)=\frac{1}{\sqrt{2}}\left(\Psi(x,t)+\Gamma\Psi^*(x,t)\right)
$$
$$
\Phi_2(x,t)=\frac{1}{\sqrt{2}i}\left(\Psi(x,t)-\Gamma\Psi^*(x,t)\right)
$$
Then each of $\Phi_1(x,t)$ and $\Phi_2(x,t)$ are a Majorana
fermion.

In spite of the beautiful simplicity of this idea, Majorana
fermions are not easy to come by in nature.  One could, for
example, decompose the relativistic electron, whose wave equation
does have a charge-conjugation symmetry, into its real and
imaginary parts. However, the interaction of the electron with
photons is not diagonal in this decomposition. The real and
imaginary components would be rapidly re-mixed by electromagnetic
interactions, they cannot be stationary states of the full
Hamiltonian of quantum electrodynamics.

One place where we might have better luck is to look for emergent
Majorana fermions in quantum condensed matter systems.  For
example, in a superconductor, the electromagnetic interactions are
effectively screened. Indeed, the Bogoliubov quasi-electrons in a
superconductor behave like neutral particles. However, even there,
in an ordinary s-wave superconductor, the anti-particle of a
quasi-electron is another quasi-electron with opposite spin.
Indeed, the quasi-electron operator in an s-wave superconductor is
the two-component object $$\left(
\begin{matrix} \psi_{\uparrow}(x) \cr \psi^*_{\downarrow}(x)\cr
\end{matrix}\right)$$ where $(\uparrow,\downarrow)$ denotes spin up
and down. It obeys the charge conjugation condition
$$\left(\begin{matrix}0&1\cr 1&0\cr\end{matrix}\right)\left(
\begin{matrix} \psi_{\uparrow}(x) \cr \psi^*_{\downarrow}(x)\cr
\end{matrix}\right)^*=\left(
\begin{matrix}
\psi_{\downarrow}(x) \cr \psi^*_{\uparrow}(x)\cr
\end{matrix}\right)$$ which is not an analog of the Majorana condition in
eqn.~(\ref{maj}), since it entails both conjugation and a flip of
the spin.

 In order to find a medium where the quasi-electron is a Majorana
fermion, we need to consider a superconductor where the condensate
has Cooper pairs with the same spin, so that the quasi-electron
has the form $$\left(
\begin{matrix} \psi_{\uparrow}(x) \cr \psi^*_{\uparrow}(x)\cr
\end{matrix}\right)$$
Then,  quasi-electron is pseudo-real, complex conjugation of its
wave-function is equivalent to multiplying by the matrix
$\Gamma=\left(\begin{matrix}0&1\cr 1&0\cr\end{matrix}\right)$,
$$\left(\begin{matrix}0&1\cr 1&0\cr\end{matrix}\right)\left(
\begin{matrix} \psi_{\uparrow}(x) \cr \psi^*_{\downarrow}(x)\cr
\end{matrix}\right)^*=\left(
\begin{matrix}
\psi_{\uparrow}(x) \cr \psi^*_{\uparrow}(x)\cr
\end{matrix}\right)$$
This gives a physical realization of a Majorana fermion. An
example of such a superconductor is one with a p-wave condensate,
such as Strontium Ruthenate~\cite{stroruth}.  There, the
condensate has the form $\langle\psi_{\uparrow}(x)\vec
x\times\vec\nabla\psi_{\uparrow}(x)\rangle$ (and can in principle
have an admixture of spin down as well). Thus, we see that, in
such a material, the quasi-electron is a two-component object
obeying a Majorana condition.   We will make use of this example
later in this Paper.

Our particular interest in the following will be in situations
where the fermion spectrum has mid-gap, or zero energy states.
These are well known to lead to interesting phenomena. Already for
complex electrons, mid-gap states give rise to fractional quantum
numbers~\cite{Jackiw:1975fn,Niemi:1984vz}. With Majorana fermions,
they are known to lead to peculiar representations of the
anti-commutator algebra which can violate basic
symmetries~\cite{Losev:2000mm,Losev:2001uc}. Some interesting
effects in the context of zero modes on cosmic strings have also
been examined~\cite{Starkman:2000bq}-\cite{Stojkovic:2000ix}.

To illustrate, let us consider the second quantization of a
complex fermion whose spectrum has a zero mode,
$$
H_0\psi_0(x)=0
$$
The conjugation symmetry implies that $$
\psi_0(x)=\Gamma\psi_0^*(x)$$

If the fermion is complex (not Majorana), the second quantized
field has a term with the zero mode wave-function and an operator,
the first term in the following expansion:
$$
\Psi(x,t)=\psi_0(x)\alpha+ \sum_{E>0}\psi_E(x)e^{-iEt/\hbar}a_E +
\sum_{E<0}\psi_E(x)e^{-iEt/\hbar}b_{-E}^\dagger~~.
$$
Here, $\alpha$ obeys the algebra
\begin{equation}
\left\{ \alpha,\alpha^\dagger\right\}=1 \label{zma}\end{equation}
 and it
anti-commutes with all of the other creation and annihilation
operators.  The existence of this zero mode leads to a degeneracy
of the fermion spectrum. The vacuum state is annihilated by all of
the annihilation operators $a_E$ and $b_{E}$. However, it must
also carry a representation of the algebra (\ref{zma}).  The
minimal representation is two-dimensional. There are two vacuum
states, $(|\uparrow>,|\downarrow>)$, which obey
$$
a_E|\uparrow>=0 = a_E|\downarrow> ~~,~~
b_{E}|\uparrow>=0=b_{E}|\downarrow>
$$
and
$$
\alpha^\dagger |\downarrow>=|\uparrow>
~~,~~\alpha^\dagger|\uparrow>=0
$$
$$
\alpha|\downarrow>=0 ~~,~~\alpha|\uparrow>=|\downarrow>
$$
The entire spectrum has a 2-fold degeneracy, with two towers of
excited states,
$$a^\dagger_{E_1}...a^\dagger_{E_m}b^\dagger_{E_1}...b^\dagger_{E_n}|\uparrow>
$$ and
$$a^\dagger_{E_1}...a^\dagger_{E_m}b^\dagger_{E_1}...b^\dagger_{E_n}|\downarrow>
$$
having the identical energies $\sum_iE_i$.

This quantization of the zero mode $\alpha$ was argued by Jackiw
and Rebbi \cite{Jackiw:1975fn} to lead to states with fractional
fermion number. Indeed, the suitably normal ordered second
quantized number operator
\begin{equation}\label{charge}
Q=\int d\vec x \frac{1}{2}\left[\psi^{\dagger}(x,t),\psi(\vec x,t)
\right]=\sum_{E>0}\left(a_E^{\dagger}a_E-b_{-E}^\dagger
b_{-E}\right)+\alpha^\dagger \alpha -\frac{1}{2}
\end{equation}
has fractional eigenvalues, for example
$$
Q|\uparrow>=+\frac{1}{2}|\uparrow>
~~,~~Q|\downarrow>=-\frac{1}{2}|\downarrow>
$$
In actuality, the charge operator is defined only up to an overall
additive constant.  However, there does exist a symmetry of the
theory, gotten at the second quantized level by replacing
$\Psi(x,t)$ by $\Gamma\Psi^*(x,t)$.  This transformation
interchanges particles and anti-particles, and is a symmetry of
the suitably normal ordered second quantized Hamiltonian.   It
should also flip the sign of $Q$.  It implies that, if there is an
eigenstate of $Q$ in the system with eigenvalue $q$,
$$Q|q>=q|q>$$
then there must exist another eigenstate $|-q>$ in the spectrum of
$Q$ with eigenvalue $-q$:
$$Q|-q>=-q|-q>$$

In addition, it is easy to argue that the eigenvalues of $Q$ are
space by integers, i.e. if $q_1$ and $q_2$ are any two
eigenvalues, then $q_1-q_2=$integer. This is essentially because
the raising and lowering operators for $Q$ are $\Psi^\dagger$ and
$\Psi$, respectively and they raise and lower in units of
integers.  In particular, this implies that
$$q-(-q)=2q={\rm integer}$$

Thus, the only possibilities are that the entire spectrum of
states have integer eigenvalues of $Q$, $q=$integer, or the entire
spectrum of states have half-odd-integer eigenvalues
$q={\small\frac{1}{2}}$-odd integer. It is easy to see that the
operator $Q$ as written in (\ref{charge}) indeed flips sign if we
interchange $a_E\leftrightarrow b_E$ and
$\alpha\leftrightarrow\alpha^\dagger$ and the offset of -1/2 that
appears explicitly there is essential for this transformation to
work.  This leads to the conclusion that, with a single fermion
zero mode, the fermion number charge is quantized in
half-odd-integer units.

Now, consider what happens for a Majorana fermion with a single
zero mode. \footnote{ We will later construct an explicit example
where this precisely this situation occurs.}  In this case, a
charge analogous to $Q$ is not defined, so the issue of fractional
charge is not relevant. But the quantization of the system is
still interesting. The second quantized operator is
$$
\Phi(x,t)= \psi_0(x)\alpha+\sum_{E>0}\psi_E(x)e^{-iEt/\hbar}a_E +
\sum_{E<0}\psi_E(x)e^{-iEt/\hbar}a_{-E}^\dagger
$$
This fermion contains half of the degrees of freedom of the
previous complex one.  Here, the $b_E$ are absent and the zero
mode operator is real, $\alpha=\alpha^\dagger$.

The creation and annihilation operator algebra is now
$$
\left\{ a_E,a_{E'}^\dagger\right\}=\delta_{EE'}
$$
as before, and
 \begin{equation}\label{majalg}
 \alpha^2=1/2
 ~~,~~\left\{\alpha,a_E\right\}=0=\left\{\alpha,a_E^\dagger\right\}
\end{equation}

A minimal representation can be constructed by defining a vacuum
state where $$a_E|0>=0~~ {\rm for~ all}~ E>0$$  Then, we can
represent the zero mode by the operator
\begin{equation}\label{zm1}
\alpha=\frac{1}{\sqrt{2}}(-1)^{\sum_{E>0}a^\dagger_Ea_E}\end{equation}
Indeed $$\alpha=\alpha^\dagger$$ and, since $$\sum_E a^\dagger_E
a_E|0>=0$$ we have $$\alpha|0>=\frac{1}{\sqrt{2}}|0>$$  The Klein
operator, $(-1)^{\sum_{E>0}a^\dagger_Ea_E}$, anti-commutes with
$a_E$ and $a_E^\dagger$. A basis for the Hilbert space consists of
the vacuum and excited states which are obtained from the vacuum
by operating creation operators
$$a_{E_1}^\dagger a_{E_2}^\dagger...a_{E_k}^\dagger|0>$$  These
are eigenstates of $\sum_{E>0}a^\dagger_Ea_E$ with integer
eigenvalues.  Thus, in this basis, $\alpha^2=1/2$ when operating
on each basis vector, and thus the identity operator on the whole
space.  The operator in (\ref{zm1}) thus satisfies the algebra
(\ref{majalg}).

Another, inequivalent representation can be obtained by starting
with
\begin{equation}\label{zm2}
\tilde\alpha=-\frac{1}{\sqrt{2}}(-1)^{\sum_{E>0}a^\dagger_Ea_E}\end{equation}
and a similar construction leads to a Hilbert space whose states
are orthogonal the one found above. We emphasize here that there
are two inequivalent representations of the anti-commutator
algebra, one where the zero mode operator is represented by
$\alpha$ in eq.~(\ref{zm1}) and one where it is represented by
$\tilde\alpha$ in eq.~(\ref{zm2}).  Both of these give an
irreducible representation and the two representations are not
related to each other by an internal automorphism.

We observe that these minimal representations of the
anti-commutator algebra have the property that they break a
symmetry of the fermion theory under $\Phi(\vec x,t)\to -\Phi(\vec
x,t)$, which we shall call ``fermion parity''. Fermion parity is a
symmetry of the linear wave equation even when $\Phi(\vec x,t)$ is
a Majorana fermion. At the quantum level, fermion parity symmetry
leads to a conservation law for the number of fermions modulo 2.
By this conservation law, any physical process must entail
creation or destruction of an even number of fermions.  For
example, if a quantum state is initially prepared with an even
number of fermions, after any physical process, the number should
remain even.   In operator language, there should exist an
operator $(-1)^F$ which anti-commutes with $\Phi(\vec x,t)$, $$
(-1)^F\Phi(\vec x,t)+\Phi(\vec x,t)(-1)^F=0
$$ and which therefore commutes with the full second quantized Hamiltonian,
$$
(-1)^FH=H(-1)^F
$$
where
$$ H=\int d\vec x \frac{1}{2}:\Phi^\dagger(\vec x,t)H_0\Phi(\vec x,t): $$

However, we see that in the minimal representations of the
anti-commutation algebras (\ref{majalg}) discussed above, in the
first representation (\ref{zm1}),
$$
<0|\Phi(x,t)|0> = + \frac{1}{\sqrt{2}}\psi_0(x)
$$
and in the second representation (\ref{zm2})
$$
<0|\Phi(x,t)|0> = - \frac{1}{\sqrt{2}}\psi_0(x)
$$
In both of these representations,  neither the vacuum state, nor
any of the excited states can be eigenstates of fermion parity,
the operator $(-1)^F$.  Thus fermion parity symmetry is broken by
the minimal quantization of this model.

Fermion parity is a sacred symmetry of physics in four dimensional
space-time~\cite{Streater:1989vi}.  All fundamental fermions in
nature have half-odd-integer spin.  A flip in sign of all fermion
operators can then be realized as a rotation by an angle $2\pi$.
Nature should be symmetric under a rotation by $2\pi$.  This means
that, if we superpose a state with even fermion number and a state
with odd fermion number, $$c_1|{\rm even}>+c_2|{\rm odd}>
$$ no experiment should be devisable, even in principle, to
measure the relative sign of $c_1$ and $c_2$.  In the four
dimensional world, unless rotation invariance is broken at a the
level of fundamental physics, we should always be free to insist
that $(-1)^F$ is a good symmetry and that we can take all physical
states as eigenstates.  Of course, this applies in four space-time
dimensions.  The emergent Majorana fermions that we want to
consider here are embedded in four space-time dimensions. We
therefore feel free to insist on fermion parity.

This brings up a contradiction with the previous discussion, where
we found that fermion parity is necessarily broken by the
quantization of the zero mode Majorana fermion system. The only way
to restore the symmetry is to use a reducible representation of the
anti-commutator algebra. The minimal modification of the
representation is equivalent to the introduction of another degree
of freedom -- and subsequent use of irreducible representations. The
new degree of freedom acts like a hidden variable.  In the
anti-commutator algebra it would be another anti-commuting variable
$\beta$ which has identical properties to $\alpha$,
$$
\beta^2=1/2$$ and anti-commutes with all other variables.  Then
the algebra of $\alpha$ and $\beta$ would have a two dimensional
representation which we could find by considering the fermionic
oscillators
\begin{eqnarray}
a=\frac{1}{\sqrt{2}}\left(\alpha+i\beta\right)~~,~~ a^\dagger =
\frac{1}{\sqrt{2}}\left( \alpha-i\beta\right) \\
\alpha=\frac{1}{\sqrt{2}}\left(a+a^\dagger\right)~~,~~ \beta =
\frac{1}{\sqrt{2}i}\left( a-a^\dagger\right) \label{zmtr}
\end{eqnarray}
which obey $$ a^2=0~~,~~a^{\dagger 2}=0~~,~~\left\{ a,
a^\dagger\right\}=1 $$ We could then find a vacuum state which is
annihilated by $a$, and another state which is created from the
vacuum by $a^\dagger$,
$$
a|->=0~~,~~a^\dagger|->=|+>
$$
$$
a|+>=|->~~,~~a^\dagger|+>=0$$ so that both are eigenstates of
$(-1)^F$ and fermion parity is restored.  Later we will see that
the hidden variable $\beta$ can have a physical interpretation.

\section{Degeneracy, tunnelling and teleportation}

In this paper, the most speculative use of Majorana fermions that
we shall find is for a kind of teleportation by quantum
tunnelling. In the context in which quantum tunnelling is normally
studied, a classical object can exist in allowed regions. There
exist other forbidden regions where it is not allowed to be. Then,
quantum tunnelling makes use of the fact that, when the particle
is quantum mechanical, its wave-function does not necessarily go
to zero in a classically forbidden region, but decays
exponentially. That means that it could, in principle, have
support on the other side of such a region and there is some small
probability that an object will be found on the other side. This
is called tunnelling.

\begin{figure}[th]
\begin{center}
\includegraphics[scale=.4]{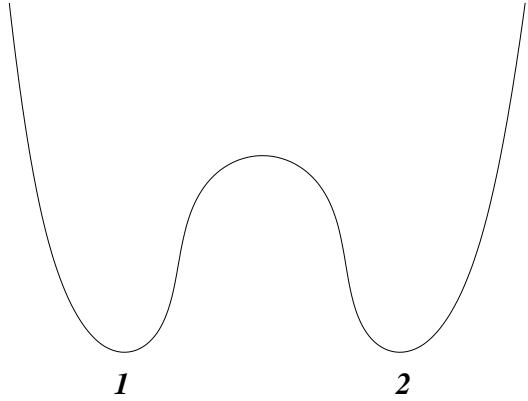}
 \caption{A double-well potential.}
\label{fig77}
\end{center}
%\vspace{8mm}
\end{figure}

\begin{figure}[th]
\begin{center}
\centerline{\psfig{figure=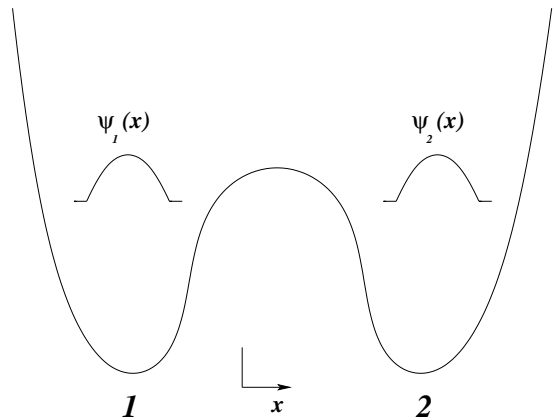,width=72mm}}
\caption{The ground state of a particle in a double well has two peaks, localized at 1
and 2.}\label{fig8}
\end{center}
%\vspace{8mm}
\end{figure}

One might try to make use of quantum tunnelling to transport an
object through a classically forbidden region.  Unfortunately, the
exponential decay of the wave-function across any classical
barrier of appreciable size renders it too small to be of any
practical use in this regard. A more sophisticated approach would
be to create a scenario where the wave-function has peaks of
appreciable size at spatially separated locations, perhaps with a
forbidden region in between.  This too will fail, but for a more
sophisticated reason which, since it is related to our later use
of Majorana fermions, we will outline. Consider, for example the
double well potential depicted in Fig. \ref{fig77}.
If the locations of
the minima are well separated and the barrier in between them is
large, semi-classical reasoning can be applied to this system.
Then, the ground state of a particle in this potential should
indeed have a peak near each of the minima, and should be
approximately symmetric under interchanging the locations of the
minima.  The typical profile of such a wave-function is drawn in
Fig. \ref{fig8}.

Now, we ask the question.  Is this state of use for tunnelling? If
this were the energy landscape in which a quantum mechanical
particle lived, could we, for example, populate this ground state
by interacting with the system in the vicinity of minimum 1 and
then depopulate the state by interacting with the system near the
other minimum - 2, effectively teleporting the particle from
location 1 to location 2?

\begin{figure}[th]
\begin{center}
\includegraphics[scale=.5]{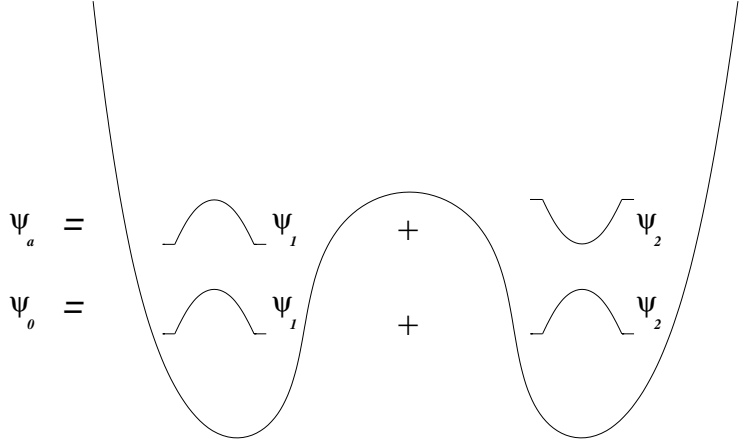}
 \caption{The almost degenerate state $\psi_a$ also has two peaks but with differing
signs.}\label{fig9}
\end{center}
%\vspace{8mm}
\end{figure}

The answer to this question is `no'. The reason for this answer is
degeneracy, or approximate degeneracy of the quantum state that we
are considering. In such a system, when our classical reasoning is
good, there must always be a second state, perhaps at slightly
higher energy but approximately degenerate with the ground state,
whose wave-function is approximately an anti-symmetric function of
the positions of the minima.  Its typical profile is depicted in
Fig. \ref{fig9}.  The ground state wave-function has the form
$\psi_0(x)=\psi_1(x)+\psi_2(x)$ where $\psi_1(x)$ is localized
near minimum 1 and $\psi_2(x)$ is localized near minimum 2.  The
anti-symmetric state would have the form
$\psi_a(x)=\psi_1(x)-\psi_2(x)$.

Now, when we interact with the system near minimum 1, while we
overlap the ground state wave-function, $\psi_0(x)$, we also
overlap $\psi_a(x)$ by the same amount. Of course, the state that
we actually populate is a linear combination of the two,
$${\small\frac{1}{\sqrt{2}}}\left(\psi_0(x)+\psi_a(x)\right)=
\sqrt{2}\psi_1(x)$$ whose wave-function is entirely localized at
the position of the first minimum.  The particle initially has
zero probability of appearing near the second minimum.   Our
attempt at teleportation by tunnelling has been foiled by
degeneracy.

Anytime the Schr\"odinger equation can be analyzed
semi-classically in this way, it seems to have a built in
protection against the long-ranged behavior that we are looking
for.

In this argument, because it is a superposition of two stationary
states with slightly differing energies, $\psi_1(x)$ is not a
stationary state. It should have a small time dependence which
eventually mixes it with $\psi_2(x)$. But this time dependence
mixes it slowly, in fact its origin is just the conventional
tunnelling amplitude for the particle to move from location 1 to
location 2 through the barrier in between.

What we need to find is a quantum system where a quantum state
which is well isolated from other states in the spectrum can have
peaks at different locations.  From the argument above, it will be
difficult to find states of this kind which obey the regular
Schr\"odinger equation.  Where we will look for such states is in
quantum condensed matter systems, where electrons, or more
properly quasi-electrons, can satisfy equations that are very
different from the Schr\"odinger equation.   To motivate this, in
the next Section, we review some of the pictorial arguments for
the appearance of fractionally charged states in polyacetylene.
Also, to set the stage for what comes next, we discuss what
happens if the fermion spectrum of polyacetylene were Majorana,
rather than complex fermions.

\section{The polyacetylene story}

\begin{figure}[th]
\begin{center}
\includegraphics[scale=.45]{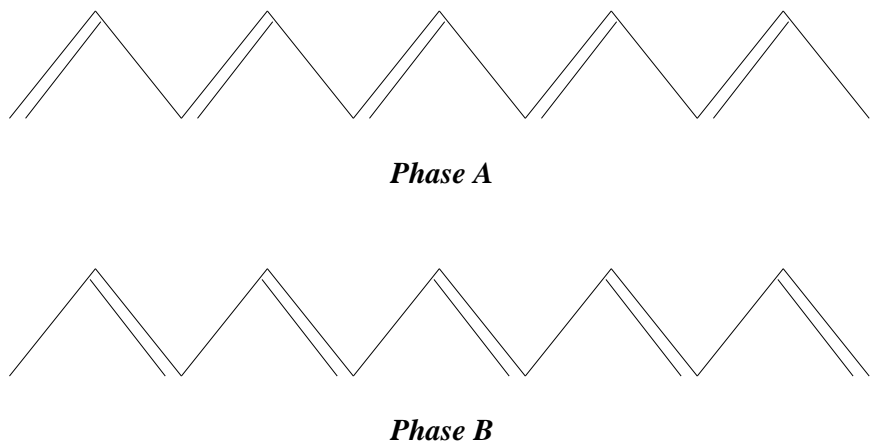}
\caption{The two degenerate ground states of polyacetylene.}
\label{fig1}
\end{center}
%\vspace{8mm}
\end{figure}

Before we consider a more quantitative model which will illustrate
our point, we pause to recall the example of the conducting
polymer, polyacetylene.  Polyacetylene is a hydrocarbon polymer
where each Carbon atom bonds with a Hydrogen atom and as well
forms two strong covalent bonds with neighboring Carbon atoms. The
fourth valence electron is nominally a conduction electron.
However, a Peirls instability localizes it into a charge density
wave which is effectively a dimer.  The result is a gap in the
electron spectrum at the fermi surface and, without impurities or
other structures, the material is an insulator. There are two
degenerate ground states, depending on the direction chosen by the
dimerization.  We illustrate these as the A and B phases in the
diagram in Fig. \ref{fig1}.
 In that figure, each line is a covalent
bond, using two of the valence electrons of the Carbon atoms.

\begin{figure}[th]
\begin{center}
\includegraphics[scale=.3]{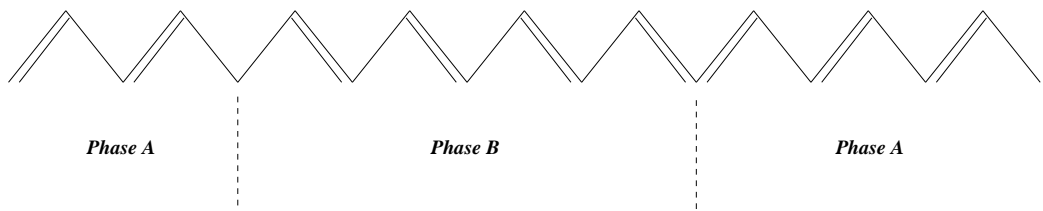}
\caption{Solitons form phase boundaries. The soliton anti-soliton pair
can be created by flipping the direction of the bonds between them.}
\label{fig2}
\end{center}
%\vspace{8mm}
\end{figure}

The conductivity of doped polyacetylene that is seen by experiments
is thought to be mostly attributed to the transport of charged
solitons along the polyacetylene molecules.   A soliton in this
system is a defect which interpolates between the two phases. We
have depicted a soliton-anti-soliton pair in Fig. \ref{fig2}.
 Note that it
can be obtained from one of the ground states by flipping the
direction of the bonds that lie between the locations of the
solitons.  Also note that the energy of the system could be higher
than that of the ground state, since the defects have non-minimal
energy configurations, but the energy density should be
concentrated in the vicinity of the solitons.  Although it will
not be an issue for us, since we are interested in other aspects
of this system, the solitons turn out to be quite mobile.  They
also carry electric charge, and can thus account for the high
conductivity that is attainable in polyacetylene.  The density of
solitons can be controlled by doping. For some original literature
on polyacetylene, see refs.~\cite{Su:1979ua}-\cite{heeger}.

\begin{figure}[th]
\begin{center}
\includegraphics[scale=.3]{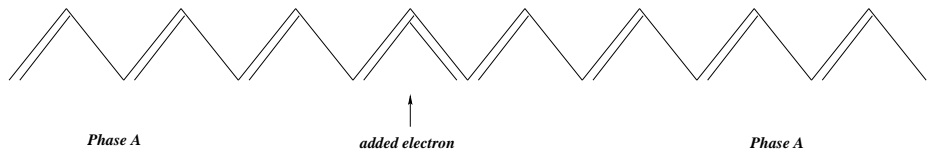}
 \caption{Phase $A$ with an additional electron.}
\label{fig3}
\end{center}
%\vspace{8mm}
\end{figure}

\begin{figure}[th]
\begin{center}
\includegraphics[scale=.3]{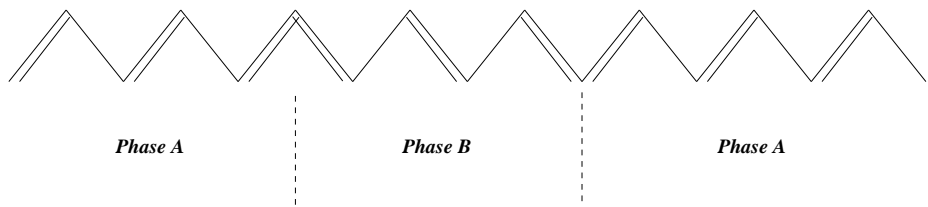}
 \caption{Beginning with phase $A$ and an additional electron, as shown in Fig. \ref{fig3},
we create a soliton-antisoliton pair which seems to share the
electron.}
\label{fig4} \end{center}
%\vspace{8mm}
\end{figure}

There is a simple argument that shows that a soliton of
polyacetylene has half of the quantum numbers of an
electron~\cite{Jackiw and Schrieffer}. In this argument, we will
neglect the spin of the electron.  Thus, for the purpose of our
arguments, in figures \ref{fig1} and \ref{fig2}, each bond stands
for a single electron, rather than a spin up, spin down pair of
electrons. Now, consider what happens when we add an electron to
phase $A$, as in Fig. \ref{fig3}.
 By flipping the directions of
some bonds, we can see that we have created a soliton-anti-soliton
pair, where each object seems to share half of the added electron.
This state is depicted in Fig. \ref{fig4}.

This brings up the question, is the electron really `split`
between the two sites?  Or does it exist in an entangled state of
some sort which has some probability -- $\small\frac{1}{2}$ -- of
the ``whole'' electron being located at either site. This question
can be made more precise by asking about measurement of the
electron charge, which is a conserved quantum number in this
system.   If, by further flipping bonds, we separate the solitons
to a large distance, and then measure the electron charge in the
vicinity of one of the solitons, is the result of the measurement
-e/2?  Or does this measurement manage to collapse the electron
wave-function somehow so that the result is either 0 or -e? In the
latter case, the average of many measurements might be -e/2, but
any single measurement would either see a whole electron or no
electron at all.  The answer to this question was found long ago
in ref.~\cite{Rajaraman:1982xf,Jackiw:1983uf}.  The conclusion was
that the measurement of the electron charge localized near one of
the solitons should yield -e/2. Put equivalently, the fractional
charge of the soliton is a sharp quantum observable. How it
manages to do this is interesting, and was discussed in
ref.~\cite{Jackiw:1983uf}.  We shall review it here.\footnote{For
some other literature on this and closely related issues, see
refs.~\cite{Ranft:1983bs}-\cite{Javanainen:2003zi}.}
 This issue has
recently been reexamined \cite{jac} in conjunction with some ideas
about entangled electron states in Helium bubbles \cite{Maris}.

\begin{figure}[th]
\begin{center}
\includegraphics[scale=.3]{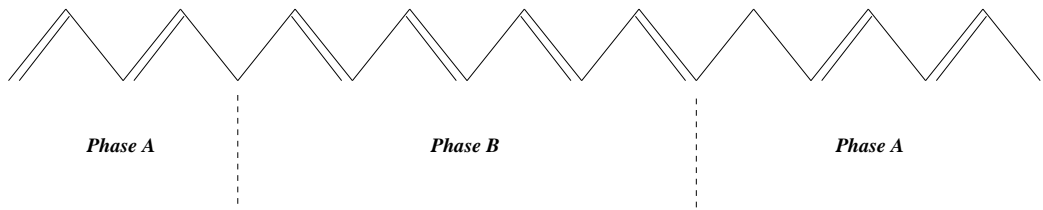}
 \caption{A soliton-antisoliton pair with a deficit of one electron.}
\label{fig5}
\end{center}
%\vspace{8mm}
\end{figure}

The electron spectrum in polyacetylene has an electron-hole
symmetry.  We could have created a state with the same energy as
the one depicted in Fig. \ref{fig4} by removing, rather than adding an
electron, to give a hole which is apparently split between the
soliton and anti-soliton, as shown in Fig. \ref{fig5}.

There are apparently four different states of the
soliton-anti-soliton system.  There are the two overall neutral
states, one of which is depicted in Fig. \ref{fig2} and the other
obtained by flipping the intermediate bonds in the opposite
direction.  We could also go from one of these states to the other
by transporting a whole electron from one soliton to the other.
The other two states we can obtain by either adding or subtracting
an electron from one of the ground states and are those that we
have already discussed in Figs. \ref{fig4} and \ref{fig5}.  We can identify
these charged soliton states in the low energy electron spectrum.  In
the single electron spectrum of polyacetylene with a
soliton-anti-soliton pair, there are two near-mid-gap states which
have small positive and negative energies. Thus the low energy
electron spectrum has four states, a ground state, an electron
state, a hole state and an electron-hole state.

\begin{figure}[th]
\begin{center}
\includegraphics[scale=.3]{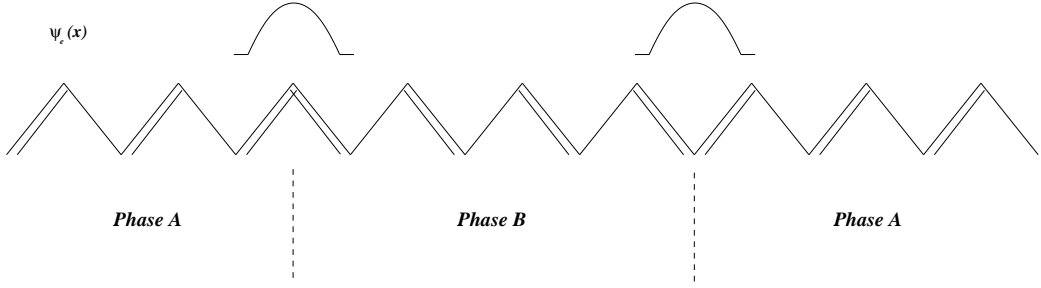}
 \caption{The electron wave-function.}
\label{fig6}
\end{center}
%\vspace{8mm}
\end{figure}

\begin{figure}[th]
\begin{center}
\includegraphics[scale=.3]{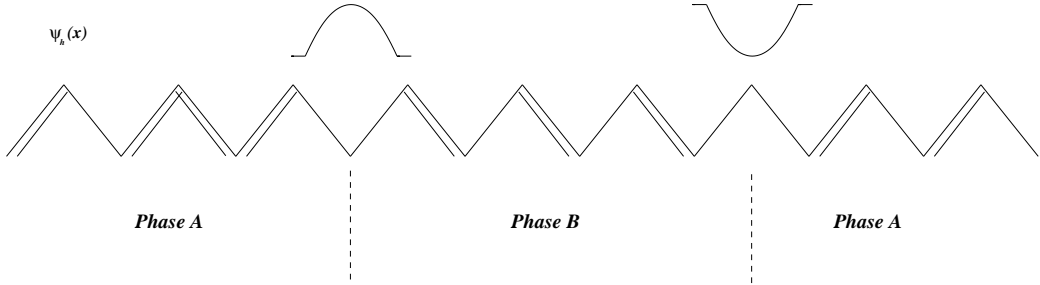}
 \caption{The hole wave-function.}
\label{fig7}
\end{center}
%\vspace{8mm}
\end{figure}

By their quantum numbers, the electron and hole states can be
identified with the configurations in Figs. \ref{fig4} and
\ref{fig5}, respectively.  The ground state and the electron-hole
state are neutral and must be formed from linear combinations of
the two neutral states. Then, in the electron state, the electron
wave-function indeed should have two peaks, as depicted in Fig.
\ref{fig6}.
Similarly the hole wave-function also should have two peaks, as is
depicted in Fig. \ref{fig7}. Detailed analysis shows that one is
an even and the other is an odd function of relative distance, as
shown in the figures.

The electron wave-function has peaks at two locations.  So, we
could ask the question again: Can we use this system for
teleportation?  Could we  populate the electron state by
interacting with one of the solitons and subsequently extract the
electron again, and thereby teleport it, by interacting with the
other soliton? To understand the answer, which will be `No!', it
is necessary to realize that, once the electric charge is a sharp
quantum observable, the electronic states of the solitons are
disentangled by a local measurement of an observable such as the
charge.

To see how this happens, let us consider a second quantization of
this system. The electron operator has the form
$$
\psi( x,t) = \psi_e(x)a +\psi_h(x)b^\dagger+...
$$
where we have identified an electron annihilation operator $a$ for
the positive energy state and a hole creation operator $b^\dagger$
for the negative energy state. We have neglected the time
dependence (the energies of the two states are exponentially small
in the soliton separation). We could as well write
\begin{equation}\label{poly}
\psi(x,t)= \psi_1(x)(a+b^\dagger) + \psi_2(x)(a-b^\dagger)+...
\end{equation}
where
$\psi_1(x)=\frac{1}{\sqrt{2}}\left(\psi_e(x)+\psi_h(x)\right)$ has
support only in the region of the left-hand soliton and $\psi_2(x)
= \frac{1}{\sqrt{2}}\left( \psi_e(x)-\psi_h(x)\right)$ has support
only near the right-hand soliton in Figs. \ref{fig6} and \ref{fig7}.

 Now, if we concentrate on the region near the
left-hand soliton, $\psi(x,t)$ or $\psi^\dagger(x,t)$ will
annihilate or create an excitation using the combination of
operators $$\alpha=\frac{1}{\sqrt{2}}\left(a
+b^\dagger\right)~~~,~~~
\alpha^\dagger=\frac{1}{\sqrt{2}}\left(a^\dagger +b\right)$$
Similarly, if we concentrate on the region around the right-hand
soliton, excitations are created and annihilated using
$$\beta=\frac{1}{\sqrt{2}}\left(a -b^\dagger\right)~~~,~~~
\beta^\dagger=\frac{1}{\sqrt{2}}\left(a^\dagger -b\right)$$ The
set of operators $(\alpha,\alpha^\dagger,\beta,\beta^\dagger)$ are
a Bogoliubov transformation of the creation and annihilation
operators $(a,a^\dagger,b,b^\dagger)$.  This transformation does
not violate fermion number -- it superposes operators with the
same fermion number.   Further, the excitations that the new
operators create or annihilate are entirely localized on one or
the other of the solitons.

Thus, again, we do not have a process whereby an electron or hole
state which has two peaks can be populated by interacting with the
system in the vicinity of one of the peaks.  We have failed to
find teleportation. Instead we have found fractional charge. The
charge density integrated over the vicinity of one of the solitons
turns out to be
$$ Q=-e\left(\alpha^\dagger\alpha-1/2\right)+{\rm ~charge~of~electrons~-
~charge~of~holes}$$
which indeed has half-odd-integer eigenvalues.  This Bogoliubov
transformation, as a mechanism for disentangling the charge
quantum numbers of the solitons was originally found in
ref.~\cite{Jackiw:1983uf}.

In eq.~(\ref{poly}), we ignored the small time dependence of the
near mid-gap states.  At this point, the reader might wonder if
the disentanglement of the soliton and anti-soliton charges that
we find by the Bogoliubov transformation would not be undone by
this time variation.  Indeed, it would be, eventually.  However
the time scale is given by the inverse of the energy gap and is
therefore exponentially large in the distance $L$ between the
soliton and anti-soliton, $T\sim m^{-1}e^{mL}$, where $m$ is the
energy gap. This is roughly the time for quantum mechanical
tunnelling between the solitons assuming an energy barrier of
height the energy gap extending over distance $L$. For macroscopic
$L$ this time $T$ should be very large.

What has prevented teleportation in this second example is again a
degeneracy, this time a slightly more subtle one since, even
though the electron and hole state have identical energies, they
have opposite signs of charge.  Avoiding teleportation has led to
fractional charge.  It has done this by a hybridization, at the
second quantized level, of the propensity of the electron field
operator to create an electron and to annihilate a hole in a local
state.

Now, imagine that, rather than complex electrons, polyacetylene
had Majorana fermions which would be obtained by identifying the
particle and hole states as the same excitations.  (Here, we are
ignoring the obvious disaster that this scenario would lead to in
chemistry.) Then, in eqn.~(\ref{poly}), we would have to identify
$a=b$ and
\begin{equation}\label{maj2}
\psi_{\rm Maj}(x,t)= \psi_1(x)(a+a^\dagger) +
\psi_2(x)(a-a^\dagger)+...
\end{equation}
Now, $a+a^\dagger$ cannot be an annihilation operator, in fact
$$(a+a^\dagger)^2=1$$ It is similar to the single zero mode
operator ``$\alpha$'' that we found for a Majorana fermion in the
Eq.~(\ref{zmtr}).  In fact, the other combination
$\frac{1}{\sqrt{2}i}(a-a^\dagger)$ now plays the role of
``$\beta$'', the ``hidden variable''. Its purpose in our previous
discussion was to provide a quantization which did not violate
fermion parity. Here, this hidden variable is just the fermion
zero mode living on the far-away anti-soliton.  We could choose
the fermion parity conserving quantization by using the states
$(|->,|+>)$ defined by
$$
a|->=0~~,~~a^\dagger|->=|+>
$$
$$
a|+>=|->~~,~~a^\dagger|+>=0
$$
which can be eigenstates of $(-1)^F$.  In these states, the
expectation value of the fermion operator vanishes, for example
$<0|\psi_{\rm Maj}(x,t)|0>=0$.  However, the two solitons are
invariably entangled.  There is now no conserved fermion number
that we can use to measure this entanglement, but there are other
effects which we will discuss in later sections once we have made
the present reasoning more solid by discussing it in the context
of a field theoretical model and them formulated a more realistic
model with emergent Majorana fermions.

\section{Relativistic Majorana fermions in a soliton background}

Single-particle states that are in some sense isolated are well
known to occur for Dirac equations, particularly when interacting
with various topologically non-trivial background fields such as
solitons, monopoles and instantons. The consequences of fermion
zero modes such as chiral anomalies \cite{Treiman:1986ep} and
fractional fermion number \cite{Jackiw:1975fn},
\cite{Niemi:1984vz} are well known.

The polyacetylene example, in the context of discussions of
fractional charge, that we used in the previous Section is a
well-known example of this.
 In polyacetylene, the low energy electron spectrum can be approximately described by
the Dirac equation~\cite{Su:1979ua,Jackiw:1982sm} and the solitons
which we discussed using pictures have a
 mathematical description as soliton-like configurations of a scalar field which
 couples to the Dirac equation.  In this Section, we will make the analysis of the
 previous Section more quantitative by considering the problem of a 1+1-dimensional relativistic
 Dirac equation coupled to a soliton background field and a soliton-anti-soliton pair.

Consider, for example, the simple one-dimensional model with Dirac
equation \begin{equation}\label{dirac} \left[ i\gamma^\mu
\partial_\mu +\phi(x)\right]\psi(x,t)=0 \end{equation}
The Dirac gamma-matrices obey the algebra $$
\left\{\gamma^\mu,\gamma^\mu\right\}=2g^{\mu\nu}$$ where
$g^{\mu\nu}=\left(\begin{matrix}1&0\cr 0&-1\cr\end{matrix}\right)$
is the (inverse of the) metric of two dimensional space-time.

This describes a fermion moving in one dimension and interacting
with a scalar field $\phi(x)$ which we shall take to have a
soliton-anti-soliton profile.  For the purposes of this
discussion, we take the ideal case of a step-function soliton
located at position $x=0$ and a step-function anti-soliton located
at $x=L$,
\begin{equation}
\phi(x)= \left\{ \begin{matrix} \phi_0 & x<0 ~,~x>L \cr -\phi_0 &
0<x<L \cr \end{matrix} \right.
\end{equation}
We will assume that the solitons are very massive, so they do not
recoil when, for example, fermions scatter from them.

 If we take $$\psi(x,t)=\psi_E(x)e^{-iEt}$$ and choose an
 appropriate basis
for the Dirac gamma-matrices, the Dirac equation becomes
\begin{equation}\label{dirachamiltonian}
i\left( \begin{matrix} 0 & \frac{d}{dx}+\phi(x)  \cr
  \frac{d}{dx}-\phi(x)& 0 \cr \end{matrix}\right) \left(
\begin{matrix} u_E(x) \cr v_E(x) \cr \end{matrix} \right) = E
\left( \begin{matrix} u_E(x) \cr v_E(x) \cr \end{matrix} \right)
\end{equation}
This equation has a particle-hole symmetry
$$
\psi_{-E}(x)=  \psi_E^*(x)
$$

It is easy to show that it has exactly two bound states. One is a
state with small positive energy and the other is the associated
hole state with a small negative energy.  The wave-functions
\begin{eqnarray}
E_+\approx +\phi_0
e^{-\phi_0L}~~~~~~~~~~~~~~~~~~~~~~~~~~~~~~~~~~~~~
~~~~~~~~~~~~~~~~~~~~~~~~~~~~
\\ \label{positiveenergystate} \psi_+(x)\approx   \sqrt{\phi_0}  \left\{
\begin{matrix}
\left(
\begin{matrix}  1 \cr  0\cr \end{matrix}
\right) e^{-\phi_0x} +{\cal O}(e^{-\phi_0L}) & x<0 \cr \left(
\begin{matrix} 1  \cr  0\cr \end{matrix}\right) e^{-\phi_0x} +
\left(
\begin{matrix} 0 \cr - i \cr \end{matrix}\right) e^{\phi_0(x-L)} +{\cal O}(e^{-\phi_0L}) & 0<x<L \cr
\left(
\begin{matrix} 0 \cr -i \cr
\end{matrix}\right) e^{\phi_0(L-x)} +{\cal O}(e^{-\phi_0L}) & L<x \cr \end{matrix}
\right.
\end{eqnarray}
\begin{eqnarray}
E_-\approx -\phi_0
e^{-\phi_0L}=-E_+~~~~~~~~~~~~~~~~~~~~~~~~~~~~~~~~~
~~~~~~~~~~~~~~~~~~~~~~~~~~~~~
\\ \label{negativeenergystate}\psi_(x)-\approx  \sqrt{\phi_0}  \left\{
\begin{matrix}
\left(
\begin{matrix}  1 \cr  0\cr \end{matrix}
\right) e^{-\phi_0x} +{\cal O}(e^{-\phi_0L}) & x<0 \cr \left(
\begin{matrix} 1  \cr  0\cr \end{matrix}\right) e^{-\phi_0x} +
\left(
\begin{matrix} 0 \cr i \cr \end{matrix}\right) e^{\phi_0(x-L)} +{\cal O}(e^{-\phi_0L}) & 0<x<L \cr
\left(
\begin{matrix} 0 \cr i \cr
\end{matrix}\right) e^{\phi_0(L-x)} +{\cal O}(e^{-\phi_0L}) & L<x \cr \end{matrix}
\right.
\end{eqnarray}
where, sufficient for our purposes, we give only the large $L$
asymptotics -- corrections to all quantities are of order
$e^{-\phi_0L}$. Note that $\psi_-(x)$ is indeed related to
$\psi_+(x)$ by $\psi_-(x)=\psi_+^*(x)$.

These states have energy well separated from the rest of the
spectrum, which is continuous and begins at $E=\pm \phi_0$. The
energies are also exponentially close to zero as the separation
$L$ is large. Furthermore, each wave-function has two peaks, one
near $x=0$ and one near $x=L$.  They have identical profile near
$x=0$ and they differ by a minus sign near $x=L$.  This is the
same feature of the electron and hole states that we claimed for
the polyacetylene soliton-anti-soliton system in the previous
Section.

The second quantized Dirac field now has the form
\begin{equation}
\psi(x,t) = \psi_+(x)e^{-iE_+t}a +
\psi^*_+(x)e^{iE_+t}b^\dagger+\ldots
\end{equation}

When $L$ is large, one can consider a second set of almost
stationary states which are the superpositions
\begin{eqnarray}\label{symmetricstate}
 \psi_0(x)=\frac{1}{\sqrt{2}}\left(e^{iE_0t}\psi_++e^{-iE_0t}\psi_-\right)
~~~~~~~~~~~~~~~~~~~~~~~~~~~~~~~~~~~~~ ~~~~~~~~~~~~~~~~~~~~~~~~~~~~
\\  \approx \sqrt{2\phi_0}  \left\{
\begin{matrix}
\left(
\begin{matrix}  \cos E_0 t \cr  0\cr \end{matrix}
\right) e^{-\phi_0x} +{\cal O}(e^{-\phi_0L}) & x<0 \cr \left(
\begin{matrix} \cos E_0t  \cr  0\cr \end{matrix}\right) e^{-\phi_0x} +
\left(
\begin{matrix} 0 \cr  \sin E_0t \cr \end{matrix}\right) e^{\phi_0(x-L)} +{\cal O}(e^{-\phi_0L}) & 0<x<L \cr
\left(
\begin{matrix} 0 \cr \sin E_0t \cr
\end{matrix}\right) e^{\phi_0(L-x)} +{\cal O}(e^{-\phi_0L}) & L<x \cr \end{matrix}
\right.
\end{eqnarray}
which has most of its support near $x=0$ and
\begin{eqnarray}\label{asymmetricstate}
 \psi_L(x)=\frac{1}{\sqrt{2}i}\left(e^{iE_0t}\psi_+ - e^{-iE_0t}\psi_-\right)
~~~~~~~~~~~~~~~~~~~~~~~~~~~~~~~~~~~~~ ~~~~~~~~~~~~~~~~~~~~~~~~~~~~
\\  \approx \sqrt{2\phi_0}  \left\{
\begin{matrix}
\left(
\begin{matrix}  \sin E_0 t \cr  0\cr \end{matrix}
\right) e^{-\phi_0x} +{\cal O}(e^{-\phi_0L}) & x<0 \cr \left(
\begin{matrix} \sin E_0t  \cr  0\cr \end{matrix}\right) e^{-\phi_0x} +
\left(
\begin{matrix} 0 \cr  -\cos E_0t \cr \end{matrix}\right) e^{\phi_0(x-L)} +{\cal O}(e^{-\phi_0L}) & 0<x<L \cr
\left(
\begin{matrix} 0 \cr -\cos E_0t \cr
\end{matrix}\right) e^{\phi_0(L-x)} +{\cal O}(e^{-\phi_0L}) & L<x \cr \end{matrix}
\right.
\end{eqnarray}
which has most of its support near $x=L$.

In terms of these wave-functions, which are localized at the sites
of the solitons,

\begin{equation}
\psi(x,t) = \psi_0(x,t){\small\frac{1}{\sqrt{2}}}\left( a +
b^\dagger\right)+ \psi_L(x,t){\small\frac{1}{\sqrt{2}i}}\left(- a
+ b^\dagger\right)+\ldots
\end{equation}
We could now consider the creation and annihilation operators
$$
\alpha={\small\frac{1}{\sqrt{2}}}\left( a + b^\dagger\right)~~,~~
\alpha^\dagger={\small\frac{1}{\sqrt{2}}}\left( a^\dagger +
b\right)
$$
$$
\beta={\small\frac{1}{\sqrt{2}i}}\left(a^\dagger - b\right)~~,~~
\beta^\dagger={\small\frac{1}{\sqrt{2}i}}\left(- a +
b^\dagger\right)
$$

By interacting with the system at $x=0$, we could as well be
dropping the fermion into the state $\psi_0$, which is localized
there and which has exponentially vanishing probability of
occurring at $x=L$ (until $\sin E_0t$ becomes appreciable, which
is just the usual estimate of tunnelling time through a barrier of
height $\phi_0$ and width $L$).

It might seem bizarre that, if we begin with the system in its
ground state when $L$ is small, then adiabatically increase $L$
that we would not simply end up with the original ground state
that has $\psi_-(x)$ populated, $\psi_+(x)$ empty. In fact, this
is a possibility. However, as we have argued in the polyacetylene
example in the previous Section,  as $L\to \infty$, the result is
an entangled state of (appropriately
defined~\cite{Rajaraman:1982xf,Jackiw:1983uf}) fermion number. If
we begin with the original ground state, measurement of the
fermion number which is localized in the vicinity of one of the
solitons will collapse the wave-function to one where the fermion,
rather than occupying the negative energy state $\psi_-$, occupies
either the state $\psi_{0}$ or the state $\psi_{L}$ which are
localized at $x=0$ or $x=L$, respectively.   As seen from the
vicinity of each soliton, these are identical to the Jackiw-Rebbi
states~\cite{Jackiw:1975fn} of the fermion in a single soliton
background, which have fermion number $\pm \frac{1}{2}$.  These
states are time-dependent, but again, just as in the polyacetylene
example, the time scale for charge fluctuations is just the
tunnelling time for a particle to go between the locations of the
solitons.

What about teleportation? Now, our dumping a fermion into the
bound state, if performed near $x=0$ would populate the state
$\psi_0(x)$, rather than $\psi_+(x)$, as all local operators would
couple only to this state. It would have appreciable probability
of appearing at $x=L$ only after a time over order
$E_0^{-1}\sim\phi_0^{-1}e^{\phi_0L}$.

The situation is somewhat different if we assume that the fermion
is a Majorana fermion. The Hamiltonian of a Majorana fermion must
have a symmetry which maps positive energy states onto negative
energy states. In the case of (\ref{dirachamiltonian}), we have $
\psi_{-E}(x)= \psi^*_{E}(x)$. Then, a fermion and an anti-fermion
have the same spectrum, and we can identify them as the same
particle.

Now, for the Majorana fermion, the pair of wave-functions
$\psi_+(x)$ and $\psi_-(x)$ correspond to the same quantum state
which can be either occupied or empty. (We can arbitrarily assign
fermion parities $(-1)^F=-1$ for the unoccupied state and
$(-1)^F=1$ for the occupied state, although $+i$ and $-i$ might be
more symmetric). In this case, the states $\psi_0$ and $\psi_L$
are wave-functions for superpositions of the occupied and
unoccupied states -- they do not have definite fermion parity.

If we begin with the system where the quantum state is an
eigenstate of fermion parity and we by some process dump a fermion
into the bound state near $x=0$, its wave-function automatically
has a second peak at $x=L$ and it could in principle be extracted
there. This defines what we mean by ``teleportation''.

If we concentrate on the region near $x=0$ and we are unaware of
the region near $x=L$, depending on the quantization, this
teleportation will appear as either violation of conservation of
fermion number mod 2 or the existence of a hidden variable in the
local theory.

\section{P-wave superconductor model and Andreev states}

Of course, the fermions in polyacetylene are not Majorana, they
are electrons with complex wave-functions. The place to look for
emergent Majorana fermions in nature is in superconductivity. Here
we shall formulate a model whose basic excitations are Majorana
fermions.  We will do this by using contact with a p-wave
superconductor to violate the conservation of total charge,
leaving behind conservation of charge modulo 2.  In such an
environment, the real and imaginary parts of the electron can have
different dynamics and the electron is essentially split into two
Majorana fermions. They can further be coupled to soliton-like
objects, in this case the boundaries of the space, in such a way
that only one of the Majorana fermions has zero modes.  Then, the
scenario that we have been looking for, an isolated
single-particle state, can be found.

In these materials, mid-gap bound states, called Andreev states,
are a common occurrence.  They typically live at surface of the
superconductor \cite{andreev}.  In our case, these will be
Majorana zero modes.

  Majorana zero modes of the type that we are
discussing are also known to be bound to vortices in p-wave
superconductors where they have the remarkable effect of giving
vortices non-Abelian fractional
statistics~\cite{stern},\cite{stone}. For concreteness we will
consider a slightly simpler model one-dimensional model that was
originally discussed by Kitaev~\cite{kitaev} in the context of
fermionic quantum computation.

\begin{figure}[ht] \hbox to \textwidth{\hfill
\begin{picture}(110,35)
\put(0,0){\line(1,0){100}} \put(0,15){\line(1,0){100}}
\put(10,35){\line(1,0){100}} \put(0,0){\line(0,1){15}}
\put(100,0){\line(0,1){15}} \put(110,20){\line(0,1){15}}
\put(0,15){\line(1,2){10}} \put(100,15){\line(1,2){10}}
\put(100,0){\line(1,2){10}}
\put(15,24.7){\thicklines\line(1,0){80}}
\put(15,25.3){\thicklines\line(1,0){80}}
\put(8,22){\footnotesize$1$} \put(96,22){\footnotesize$L$}
\put(0,4){\hbox to 100\unitlength
{\footnotesize\hfil${\rm p-wave~sc}$\hfil}}
\end{picture}
\hfill} \caption{A  quantum wire embedded in a bulk P-wave
superconductor.} \label{fig_layout}
\end{figure}
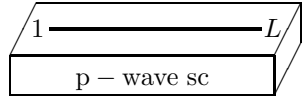

We shall consider a quantum wire embedded in a bulk P-wave
superconductor as is depicted in Fig. \ref{fig_layout}.

We shall assume that the wire has a single channel.  We shall also
assume that the dynamics of electrons in the wire are adequately
described by a one-dimensional tight-binding model.  We will
ignore the spin degree of freedom of the electron.  The phenomenon
that we will find is to a first approximation spin-independent.

We will assume that the coupling to the neighboring p-wave
superconductor is weak and its net effect is to give electrons the
possibility of entering and leaving the wire in pairs by creating
or destroying a p-wave cooper pair in the bulk. To describe the
electrons, we will use the Hamiltonian
\begin{equation}\label{mbham}
H=\sum_{n=1}^L \left( \frac{t}{2}a_{n+1}^\dagger
a_n+\frac{t^*}{2}a_n^\dagger a_{n+1}+\frac{\Delta}{2}
a^{\dagger}_{n+1}a^{\dagger}_n + \frac{\Delta^*}{2} a_n
a_{n+1}+\mu a_n^\dagger a_n\right)
\end{equation}
Sites on the quantum wire are labelled by $n=1,2,...,L$.  The
operators $a_n$ and $a_n^\dagger$ annihilate and create an
electron at site $n$. They obey the anti-commutator algebra
\begin{equation}\label{latticeacr} \left\{ a_n, a_{n'}^\dagger\right\}=\delta_{nn'}
\end{equation}

The first terms in the Hamiltonian, with coefficients $t$ and
$t^*$ are the contribution to the energy of the hopping of
electrons between neighboring sites.
 The second pair of terms, with $\Delta$ and $\Delta^*$, arise
from the presence of the super-conducting environment. They
describe the amplitude for a pair of electrons to leave or enter
the wire from the environment.   It is assumed that they can do
this as a Cooper pair when they are located on neighboring sites.
This is effectively an assumption about the size and coherence of
the cooper pairs in the superconductor.  Even if it were not
accurate, the smaller next-to-nearest neighbor, etc. terms that
would arise could be taken into account and would not change our
result significantly.  The last term is the chemical potential,
the energy of an electron sitting on a site of the wire. We shall
assume a reasonable hierarchy of the parameters, that the
amplitude for hopping along the wire is somewhat larger than
hopping to and from the bulk, $|t|>|\Delta|$, and that the
chemical potential is close enough to zero that the electron band
has substantial filling, $|\mu|<|t|$.

\subsection{Spectrum of single-particle states}

Let us discuss the spectrum of the single-particle states in the
many-body theory described by the Hamiltonian (\ref{mbham}). If
$t=|t|e^{i\phi}$ and $\Delta = |\Delta|e^{2i\theta}$, by
redefining $a_n\to e^{i(\phi+\theta)}a_n$ for $n$ odd and $a_n\to
e^{i(\phi-\theta)}a_n$ for $n$ even, we remove the complex phases
of $t$ and $\Delta$, which we can henceforth assume to be positive
real numbers. The equation of motion for the fermion wave-function
is gotten by taking the commutator of its operator $a_k$ with the
Hamiltonian (\ref{mbham}),
$$
i\hbar\dot a_n = \left[ a_n, H\right]
$$
for which we get \begin{equation}
 i\hbar\frac{d}{dt} a_n =
\frac{t}{2}\left(a_{n+1}+a_{n-1}\right)-\frac{\Delta}{2}\left(a_{n+1}^\dagger
- a_{n-1}^\dagger\right) + \mu a_n \label{wave}\end{equation} for
the sites $n=2,...,L-1$.

Because we are using open boundary conditions -- the chain simply
ends at $n=1$ and $n=L$, the equations for $\frac{d}{dt} a_1$ and
$\frac{d}{dt} a_L$ differ from (\ref{wave}) by missing terms. When
we solve (\ref{wave}) as a wave equation, it will be convenient to
deal with this by extending the chain by one site in each
direction and then eliminating the extra sites by imposing
Dirichlet boundary conditions,
$$
a_0(t)=0~~,~~a_{L+1}(t)=0
$$
With these conditions, (\ref{wave}) describes the dynamics for all
$n=1,2,...,L$.

 Now, it is most efficient to decompose the electron into real and imaginary
parts, $a_n=b_n+ic_n$, and assemble them into a spinor
\begin{equation}\label{spinor}
\psi_n = \left(
\begin{matrix} b_n \cr c_n \cr \end{matrix} \right) \end{equation}
Note that this spinor obeys the Majorana condition
\begin{equation}\label{cond}\psi_n = \psi_n^* \end{equation}

The equation for the wave-function is
\begin{equation}
\left(\begin{matrix} \mu & \hbar\frac{d}{dt}\cr -\hbar\frac{d}{dt}
& \mu\cr\end{matrix}\right)\psi_n+ \left(\begin{matrix}
\frac{1}{2}(t-\Delta)&0\cr 0&\frac{1}{2}(t+\Delta)\cr
 \end{matrix}\right)\psi_{n+1}  + \left(\begin{matrix}
\frac{1}{2}(t+\Delta)&0\cr 0&\frac{1}{2}(t-\Delta)\cr
 \end{matrix}\right)\psi_{n-1}=0
 \label{diff}
 \end{equation}

In order to solve the equation, we will make the ansatz
\begin{equation}\label{fourier}
\psi_n(t) = e^{-i\omega t/\hbar}\psi_n(\omega)
\end{equation}
The Majorana condition for energy eigenstates is
$$\psi_n^*(\omega)=\psi_n(-\omega)$$  We will  normalize
the wave-functions with the condition
$$
\sum_{n=1}^L\left|\psi_n(\omega)\right|^2~=~1~=~\sum_{n=1}^L
\left( |b_n(\omega)|^2+|c_n(\omega)|^2\right) $$

Since the equation and boundary conditions are linear, we can
further make the ansatz that the wave-functions are superpositions
of plane waves,
\begin{equation}
\psi_n(\omega) = \zeta^n \left( \begin{matrix} u(\zeta)\cr
v(\zeta)\cr\end{matrix}\right)
\end{equation}
Then, the difference equation (\ref{diff}) becomes

The equation for the wave-function is
\begin{equation}\label{matrix}
\left(\begin{matrix}
\frac{1}{2}t(\zeta+1/\zeta)-\frac{1}{2}\Delta(\zeta-1/\zeta)+\mu &
-i\omega \cr i\omega &
\frac{1}{2}t(\zeta+1/\zeta)-\frac{1}{2}\Delta(\zeta-1/\zeta)+
\mu\cr\end{matrix}\right) \left( \begin{matrix} u(\zeta)\cr
v(\zeta)\cr\end{matrix}\right)=0
 \end{equation}
 which has a solution when the frequencies obey the dispersion
 relation
 \begin{equation}\label{dispersion}
 \omega^2 = \left[ \frac{1}{2}t(\zeta+1/\zeta)+\mu\right]^2
 -\left[\frac{1}{2}\Delta(\zeta - 1/\zeta)\right]^2
 \end{equation}
 For a given real value of $\omega$, there are generally four
 wave-vectors which satisfy this dispersion relation,
 $$
 \zeta_\omega~,~1/\zeta_\omega~,~ \zeta^*_\omega~,~1/\zeta^*_\omega
$$
To find a solution of the wave equation, we must take
superpositions of the four solutions of (\ref{matrix}) with each
of these four wave-vectors.  Then we must adjust the four
coefficients of the superposition in order to satisfy the four
boundary conditions.  (Remember that the boundary conditions are
for spinors, so there are four boundary conditions in total.)
Three of the boundary conditions can be solved by adjusting the
coefficients in the superposition. The fourth superposition
coefficient can eventually be determined up to phases by
normalizing the wave-function.   The fourth boundary condition,
which has yet to be satisfied, then gives a condition that the
wave-vector must obey. Plugging the resulting wave-vector back
into the dispersion relation (\ref{dispersion}) then gives the
allowed energy eigenvalue.
 This gives an algorithm for finding the energies, the allowed
 wave-vectors (which are $\frac{1}{i}\ln\zeta$ and are generally complex)
 and the wave-functions.

When $L$ is large, the solutions are of two kinds.  One are to a
good approximation continuum states, where $\zeta=e^{ik}$ and the
continuum spectrum is
$$
\omega(k)=\pm\sqrt{ \left[t\cos k+\mu\right]^2+\Delta^2\sin^2k}
$$
with $k\in (-\pi,\pi]$ (it is quantized approximately as
$k=2\pi\cdot{\rm integer}/(L+1)$ which becomes a continuum when
$L\to\infty$).  This spectrum has an energy gap.  The point of
closest approach of the positive and negative energy bands occurs
when $\cos k=-t\mu/(t^2-\Delta^2)$ and the gap is $E_{\rm
gap}=2\Delta\sqrt{ \frac{t^2-\Delta^2-\mu^2}{t^2-\Delta^2}}$.  We
will assume that this gap is significant, so that the mid-gap
states that we will discuss next are indeed well isolated.

The other modes in the spectrum are a pair of mid-gap states.
When $L$ is large, these states have energies that are
exponentially small in $L$, one is positive, one is negative and
they have equal magnitudes. In the following, we will solve for
the spectrum of these mid-gap states in the approximation where
effects that are exponentially small in $L$ are neglected.

 We begin with an un-normalized spinor
 \begin{eqnarray}
 \zeta^n\left( \begin{matrix} i\omega\cr
\frac{1}{2}t(\zeta+1/\zeta)-\frac{1}{2}\Delta(\zeta-1/\zeta)+\mu
\cr\end{matrix}\right)+A\zeta^{-n}\left( \begin{matrix} i\omega\cr
\frac{1}{2}t(\zeta+1/\zeta)+\frac{1}{2}\Delta(\zeta-1/\zeta)+\mu
\cr\end{matrix}\right) \nonumber \\
+B\zeta^{*n}\left(
\begin{matrix} i\omega\cr
\frac{1}{2}t(\zeta^*+1/\zeta^*)-\frac{1}{2}\Delta(\zeta^*-1/\zeta^*)+\mu
\cr\end{matrix}\right)+C\zeta^{*-n}\left( \begin{matrix}
i\omega\cr
\frac{1}{2}t(\zeta^*+1/\zeta^*)+\frac{1}{2}\Delta(\zeta^*-1/\zeta^*)+\mu
\cr\end{matrix}\right) \label{solution}\end{eqnarray}  We will
solve the boundary condition for the mid-gap state in the limit
where $L$ is large.  There, we expect the solution to be very
close to $\omega=0$, for which we then need a wave-vector which
solves $t(\zeta+1/\zeta)+2\mu=-\Delta(\zeta-1/\zeta)$.   Then, to
a first approximation, the terms with $A$ and $C$ are absent from
(\ref{solution}) and we must choose the
$B=-(\zeta-1/\zeta)/(\zeta^*-1/\zeta^*)$ in order to satisfy the
boundary condition at $n=0$.  Since $$
\zeta=-\frac{\mu}{2(t+\Delta)}+
i\sqrt{\frac{t-\Delta}{t+\Delta}}\sqrt{1-\mu^2/4(t^2-\Delta^2)}$$
so that $\zeta\zeta^*=\frac{t-\Delta}{t+\Delta}<1$, this gives a
wave-function which is maximal at $n=1$ and which decays
exponentially as $n$ increases.  This would indeed be the solution
for the mid-gap state on the half-line when $L\to \infty$. When
$L$ is finite, rather than infinite, in order to satisfy the
boundary condition at $n=L+1$ we must include an amplitude for the
growing solution.  It can be obtained from the decaying one by
simply replacing $n$ by $L+1-n$ and multiplying the spinor by
$\sigma^2$. Thus, to a good approximation the mid-gap solution is
\begin{eqnarray}
\psi^+_n=\sqrt{
\frac{\Delta}{2t}\frac{t^2-\mu^2}{t^2-\Delta^2-\mu^2}}\left[
\frac{
 \left( -\mu+i\sqrt{t^2-\Delta^2-\mu^2}\right)^n-\left(
-\mu-i\sqrt{t^2-\Delta^2-\mu^2}\right)^n } {(t+\Delta)^n
}\left(\begin{matrix}0\cr i\cr \end{matrix}\right)+\right.\nonumber \\
\left.  +\frac{  \left(
-\mu+i\sqrt{t^2-\Delta^2-\mu^2}\right)^{L+1-n}-\left(
-\mu-i\sqrt{t^2-\Delta^2-\mu^2}\right)^{L+1-n} }
{(t+\Delta)^{L+1-n} }\right]\left(\begin{matrix}1\cr 0\cr
\end{matrix}\right)\nonumber
\end{eqnarray}
This wave-function has infinitesimal positive energy.  The
wave-function with infinitesimal negative energy is given by
\begin{eqnarray}
\psi^-_n=\sqrt{
\frac{\Delta}{2t}\frac{t^2-\mu^2}{t^2-\Delta^2-\mu^2}}\left[
\frac{
 \left( -\mu+i\sqrt{t^2-\Delta^2-\mu^2}\right)^n-\left(
-\mu-i\sqrt{t^2-\Delta^2-\mu^2}\right)^n } {(t+\Delta)^n
}\left(\begin{matrix}0\cr i\cr \end{matrix}\right)-\right.\nonumber \\
\left.  -\frac{  \left(
-\mu+i\sqrt{t^2-\Delta^2-\mu^2}\right)^{L+1-n}-\left(
-\mu-i\sqrt{t^2-\Delta^2-\mu^2}\right)^{L+1-n} }
{(t+\Delta)^{L+1-n} }\right]\left(\begin{matrix}1\cr 0\cr
\end{matrix}\right)\nonumber
\end{eqnarray}
We will abbreviate these by naming the function
\begin{equation}\label{phin}\phi_n=i\sqrt{
\frac{\Delta}{2t}\frac{t^2-\mu^2}{t^2-\Delta^2-\mu^2}} \frac{
 \left( -\mu+i\sqrt{t^2-\Delta^2-\mu^2}\right)^n-\left(
-\mu-i\sqrt{t^2-\Delta^2-\mu^2}\right)^n } {(t+\Delta)^n }
\end{equation} where $\phi_n=\phi_n^*$ and we have normalized to \begin{equation} \sum_n
|\phi_n|^2=\frac{1}{2}\end{equation} The function $\phi_n$ has
maximum magnitude at $n=1$ and it decays exponentially as $n$
increases.   We shall use the notation
\begin{eqnarray}
\psi^+_n=\phi_n\left(\begin{matrix}0\cr 1\cr \end{matrix}\right)-
\phi_{L+1-n}\left(\begin{matrix}i\cr 0\cr
\end{matrix}\right)
\end{eqnarray}

\begin{eqnarray}
\psi^-_n=\phi_n\left(\begin{matrix}0\cr 1\cr
\end{matrix}\right)+\phi_{L+1-n}\left(\begin{matrix}i\cr 0\cr
\end{matrix}\right)
\end{eqnarray}
We have normalized the spinors so that
\begin{equation}\sum_n\psi^{\pm\dagger}_n\psi^\pm_n=1\end{equation}

Note that, these wave-functions satisfy the Majorana condition
$\psi^-_n=\psi^{+*}_n$.  As expected, they have support near $n=1$
and $n=L$ and are exponentially small in the interior of the
quantum wire, far from the boundaries. Further, we have adjusted
phases so that the wave-functions are identical in profile in the
region near $n=1$. Then, we expect that they differ in sign in the
region near $n=L$ and we confirm from that above that this is so.
Also, note that they are complex. To form the real, Majorana
spinor, we must superpose them with a creation and annihilation
operator,
\begin{equation}
\psi_n(t)= \psi^+_n e^{-i\omega t}a+\psi^-_n e^{i\omega
t}a^\dagger + {\rm ~non-zero~energy~states}
\end{equation}
Here $a$ and $a^\dagger$ are the annihilation and creation
operators for the mid-gap state and $\omega$ is their
exponentially small energy. Ignoring the energy, we can also write
this operator as
\begin{eqnarray}
\psi_n(t)= \phi_n\left(\begin{matrix}0\cr 1\cr
\end{matrix}\right)(a+a^\dagger)+\phi_{L+1-n} \left(\begin{matrix}1\cr 0\cr
\end{matrix}\right)\frac{1}{i}(a-a^\dagger)+\ldots\nonumber
\end{eqnarray}
The first term on the right-hand side has support near $n=0$ and
decays exponentially as $n$ increases from $1$.  The second term
has support near $n=L$ and decays exponentially as $n$ decreases
from $L$. They each multiply the operators
$\alpha=\frac{1}{\sqrt{2}}\left((a+a^{\dagger}\right)$ and
$\beta=\frac{1}{\sqrt{2}i}\left(a-a^{\dagger}\right)$,
respectively. These are analogous to the operators which we
introduced on Section 1. $a$ and $a^\dagger$ must have the
anti-commutator
$$
\left\{ a,a^\dagger\right\}=1
$$
which has a two-dimensional representation, the states $|->$ and
$|+>$ of Section 1 which we copy here for the reader's
convenience, $$ a|->=0~~,~~a^\dagger |->=|+>$$ $$ a|+>=|->~~,~~
a^\dagger |+>=0 $$
 All other excited states of the system are created
by operating creation operators for the other, non-zero energy
excitations.  Remember that it is the states $|+>$ and $|->$ which
we expect to be eigenstates of fermion parity, $(-1)^F$.

\subsection{Second quantized electron operator}

Now, we recall that the upper and lower components of the spinor
$\psi_n(t)$ that we discussed in the previous subsection are
simply the real and imaginary components of the electron field
operator, which we can now reconstruct,
\begin{eqnarray}\label{electronfieldoperator}
a_n(t)=  \phi_{L+1-n}\frac{1}{i}\left(a-a^\dagger\right)+ i\phi_n
\left(a+a^{\dagger}\right)+\ldots
\end{eqnarray}
This is now a complex operator, but its real and imaginary parts
have support at opposite ends of the quantum wire.  The part of
the operator which has not been written, and is indicated by dots
in (\ref{electronfieldoperator}), are superpositions of creation
and annihilation operators for continuum states.  All such states
have energies above the gap and extended, plane-wave-like
wave-functions.   Note that now that the phase symmetry of the
system has been broken by coupling to the superconductor, the real
and imaginary parts of the electron operator will generally have
different properties.  This interesting fact will not concern us
in the following and we will focus on the mid-gap, or zero mode
part of the electron operator.

Note, now, if we operate with any local operator in the vicinity
of $n=1$, the electon operator acts as if it were composed of the
combination of zero mode operators $(a+a^\dagger)$.  As we have
discussed before, this operator squares to a constant.  There
cannot be any states that it annihilates.  Thus, operating it on
any state of the system, in the region where the zero mode
wavefunction has support, will have an effect.  What it does is
flip the state from $|->$ to $|+>$.  Since it is a hermitian
operator, it is possible to diagonalize it, the states
$\frac{1}{\sqrt{2}}(|->+|+>)$ and $\frac{1}{\sqrt{2}}(|->+|+>)$
are its eigenvectors.   However, these eigenvectors are not
eigenstates of fermion parity.

\section{Long ranged correlations of electrons}

What about teleportation?  Let us imagine that we begin with the
system in one of its ground states, say $|->$ and inject an
electron so that at time $T=0$ it is resting at site $\# 1$.  This
means, we being with the state $a_1^{\dagger}|->$, where, as we
recall, $a_1^\dagger$ is the creation operator for an electron at
site $\#$1.

We then ask what is the quantum transition amplitude for the
transition, after a time $T$ has elapsed, of this state to one
with the electron located at position $\# L$.  The final quantum
state is $a_L^{\dagger}|->$. The amplitude is given by
\begin{equation} {\cal A} = <-|~a_L ~e^{iHT}~
a_1^{\dagger}~|-> = |\phi_1^0|^2+\left(T{\rm ~ and~}
L{\rm-dependent} \right)
\end{equation}
The $T$- and $L$-dependent parts of this matrix element represent
the usual propagation via excited quasi-electrons which must
travel across the wire. The first term is non-zero and is $T$ and
$L$-independent. By `teleportation', we are referring to this part
of the amplitude. Here, we can evaluate the amplitude explicitly.
It is \begin{equation} {\cal A}_{\rm
Tel}=\left(\frac{2\Delta}{t}\right) \left( \frac{
t^2-\Delta^2-\mu^2}{(t+\Delta)^2}\right)
\end{equation}
Which can be appreciable, in the 10-30 percent range, for a
surprisingly wide choice of parameters.

However, the teleportation probability is the square of this
amplitude, which is somewhat smaller.   We could ask a more
sophisticated question: What is the probability that the electron,
once injected at $n=1$ appears anywhere within the exponential
range of the zero mode wave-function at $n=L$.  This probability
would be given by
\begin{equation}
{\cal P}_{\rm Tel} = \sum_n |\phi_n|^2 |\phi_1^0|^2
=\frac{1}{2}\left(\frac{2\Delta}{t}\right) \left( \frac{
t^2-\Delta^2-\mu^2}{(t+\Delta)^2}\right)
\end{equation}
This is what we shall call the ``teleportation probability``.
Again, for a range of parameters $t,\Delta$ and $\mu$, it can be
appreciable.

\section{Discussion}

The apparently instantaneous propagation of an electron would seem
to be a potential violation of Einstein causality, since in
principle a message could be sent at a speed faster than that of
light.

Let us review the nature of the system that we have constructed.
Once the quantum wire - p-wave superconductor system is prepared,
the extended Majorana state of the electron is already there,
ready for use. The system has a two-fold degeneracy: at low
energy, there are two states $|->$ and $|+>$. These are not normal
quantum states in that they differ by a quantum number which we
would like to preserve, fermion parity $(-1)^F$.

Thus, if we do not allow superpositions of these states, this is
effectively a classical bit, like a classical switch that can
either be OFF or ON, the wave-function can be in one state or the
other.

The system moves from OFF to ON by absorbing or emitting an
electron in a way that flips the vacuum from one state to the
other.  This should occur somewhere in the vicinity of the ends of
the wire, where the zero mode wave-functions have support. It can
move back from ON to OFF by the identical process, again absorbing
or emitting a single electron.

This leads to the rather drastic conclusion that there could be
super-luminal transfer of information in this system. One would
need only to prepare the system in one of its ground states, with
a sender sitting at 1 and a receiver sitting at L.  Either ground
state is sufficient and neither the sender nor the observer needs
to know which it is.  All the receiver has to do is wait for an
electron to arrive. If it arrives with energy at or above the
electron energy gap, he or she can conclude that it propagated
normally and was sent at some time in the past.  However, if it
arrives at very low energy, he or she knows that it tunnelled and
that it was sent by the sender at that instant.  This is seemingly
an instant transfer of information over a finite distance.

There is a obvious way out of this, but it means giving up the
fermion parity symmetry that has until now been sacred.  If we
allow superpositions  of the states $|->$ and $|+>$ which have
even and odd fermion number, then the degenerate ground states are
a quantum rather than classical two-level system, there are two
states and any superpositions are allowed. Now, in this system, it
is easy to prepare states where an electron can spontaneously
appear or disappear. Take, for example an eigenstate of the
operators that we called $\alpha$ and $\beta$.  In their
eigenstates, $\frac{1}{\sqrt{2}}\left(|->\pm|+>\right)$, the
electron operator has an expectation value
$<a_n(t)>=\pm\phi_n\pm\phi_{L+1-n}$.  It would thus have an
amplitude for simply vanishing or appearing spontaneously.

Then, when the observer at L detects the arrival of a low energy
electron he or she cannot distinguish one which was sent from the
other side of the wire from one which is spontaneously created.
This restores Einstein causality at the expense of our having to
admit states onto physics which are not eigenstates of fermion
parity. There is the further question of whether such states are
consistent with three dimensional physics.

Fermion number mod 2 is an important conservation law in three
dimensional physics~\cite{Andreev:2003nf}.  Even though the
quantum wire that we have discussed is one-dimensional, it is
embedded in three dimensional space and the electrons that we are
discussing are spin-$\frac{1}{2}$ particles in three dimensional
space. This means that their wave-functions individually change
sign under a rotation by $2\pi$. More importantly, a state with
odd fermion number must change sign under a rotation by $2\pi$
whereas a state with even number should remain unchanged. A
rotation by angle $2\pi$ should not affect physics. Thus, the
relative sign of even and odd fermion number states should not
have any physical consequences.

If we did allow a superposition of the two states, they would form
a single qubit.  We could parameterize the state-vector by a point
on the Bloch sphere $(\theta,\phi)$ where the state is
\begin{equation}
|\theta,\varphi>~=~\cos{\small\frac{\theta}{2}}|->+e^{i\varphi}\sin{\small\frac{\theta}{2}}
|+>
\end{equation}
Points on the two-dimensional unit sphere are specified by the
unit vectors
$$\hat n =(\sin\theta\cos\varphi,\sin\theta\sin\varphi,\cos\theta)$$ and $0\leq
\theta\leq\pi$, $-\pi<\varphi\leq\pi$. However, as we have argued,
the relative sign of the two states should not be an observable.
Then the set of ``physical states'' of the qubit would be the
Bloch sphere with a further identification
\begin{equation}\label{identification}
\varphi \sim \varphi+\pi
\end{equation}
Of course, this identification is allowed only if there are no
experiments, even in principle, which could measure the relative
sign of the two states in the superposition.   Normally, one could
measure that sign by an interference experiment.

For example, we could attempt to observe the relative sign by
examining interference between the electron which arrives by
tunnelling and the one which arrives by conventional transport.
However, the teleportation amplitude in the state $\phi$
\begin{equation}  <\theta,\varphi|~a_L ~e^{iHT}~
a_1^{\dagger}~|\theta,\varphi>=\cos\theta[{\rm
teleportation}]+[{\rm transport}]
\end{equation}
The teleportation amplitude is diminished by a factor of
$\cos\theta$ whereas the transport amplitude is unchanged.  One
can make the teleportation amplitude vanish by adjusting
$\theta=\pi/2$.  However, the relative amplitude cannot be used to
measure the relative sign of the two components of the
wave-function.

There is an amplitude for an electron to vanish,
\begin{equation}  <\theta,\varphi|~e^{iHT}~
a_1^{\dagger}~|\theta,\varphi>\sim
i\sin\theta\cos\varphi~\cdot\phi_1
\end{equation}
and to appear spontaneously
\begin{equation}  <\theta,\varphi|~a_L ~e^{iHT}~
 ~|\theta,\varphi>\sim \sin\theta\sin\varphi~\cdot\phi_1
\end{equation}
As we expect, the latter two amplitudes change sign when we put
$\varphi\to\varphi+\pi$. Actual quantum observables are
probabilities which are the modulus squares of amplitudes.   They
are also insensitive to the relative sign of the two parts of the
wave-function.

The above probability amplitudes do not offer a way to distinguish
the quantum states with $\varphi$ and $\varphi+\pi$.  At this
point, we have not ruled out, but also we have not devised an
experiment by which they could be distinguished.  Indeed, if there
is no such experiment, we are free to cut the Bloch sphere in half
by the identification (\ref{identification}) and the ground states
would form this peculiar qubit.   Teleportation still happens, but
so does the spontaneous disappearance or appearance of a single
electron and the contradiction with Einstein locality is removed.

 We cannot exclude the possibility that the effect that we have
been discussing could be interfered with by the superconductor
which the quantum wire is in contact with.   Here, we have assumed
that it acts as a simple bath which supplies and absorbs Cooper
pairs but is otherwise innocuous. We cannot rule out that it also
has exotic states that should be included in the picture.

\section*{Acknowledgements}
G.S. thanks  Duncan Haldane, Tony Leggett, Girard Milburn and Bill
Unruh for discussions and acknowledges the hospitality of the
I.H.E.S., Bures-sur-Yvette, the Werner Heisenberg Institute,
Munich and the University of Perugia where parts of this work were
completed. P.S. thanks M.Rasetti, P. Zanardi and V. Barone for
discussions and acknowledges the Pacific Institute for Theoretical
Physics, Vancouver and the Center for Theoretical Physics at MIT
for hospitality. This work is supported in part by NSERC of Canada
and by M.I.U.R.~National Project ''JOSNET'' (grant n.2004027555).

\end{document}